# Imaging polar and dipolar sources of geophysical anomalies by probability tomography. Part I: theory and synthetic examples

**Paolo Mauriello** [1] and **Domenico Patella** [2]

[1] *Department of Science and Technology for Environment and Territory, University of Molise, Campobasso, Italy*
*(E-mail: mauriello@unimol.it)*

[2] *Department of Physical Sciences, University Federico II, Naples, Italy*
*(E-mail: patella@na.infn.it)*

**ABSTRACT**

We develop the theory of a generalized probability tomography method to image source poles and dipoles of a geophysical vector or scalar field dataset. The purpose of the new generalized method is to improve the resolution power of buried geophysical targets, using probability as a suitable paradigm allowing all possible equivalent solution to be included into a unique 3D tomography image. The new method is described by first assuming that any geophysical field dataset can be hypothesized to be caused by a discrete number of source poles and dipoles. Then, the theoretical derivation of the source pole occurrence probability (SPOP) tomography, previously published in detail for single geophysical methods, is symbolically restated in the most general way. Finally, the theoretical derivation of the source dipole occurrence probability (SDOP) tomography is given following a formal development similar to that of the SPOP tomography. The discussion of a few examples allows us to demonstrate that the combined application of the SPOP and SDOP tomographies can provide the best core-and-boundary resolution of the most probable buried sources of the anomalies detected within a datum domain.

## INTRODUCTION

Probability tomography has recently been proposed in applied geophysics as a method to virtually explore the subsoil in the search of the most probable emplacement of the sources of anomalies appearing in a field dataset collected on the ground surface. It has originally been formulated for the self-potential method (Patella, 1997 a, b). Afterward, it has been extended to the geoelectric (Mauriello *et al*., 1998; Mauriello and Patella, 1999 a, 2005 b), *em* induction (Mauriello and Patella, 1999 b, 2000), gravity (Mauriello and Patella, 2001 a, b) and magnetic methods (Mauriello and Patella, 2005 a).

In this paper we generalize the theory of probability tomography to any geophysical anomalous field, under the assumption that the dataset, collected in a generic datum domain (volume, surface or line), can be viewed as the response of a double set of hidden sources, say poles and dipoles, requiring to be identified.

The poles have been the only sources postulated in the original treatment. They have been assumed as the most elementary sources of physical nature, capable of reproducing, with appropriate spatial combinations, the responses due to macroscopic entities. No recourse to an a priori choice of a class of models, able to generate synthetic responses compatible with the field ones as in any standard deterministic interpretation approach, has been included in the probability tomography approach, thus avoiding any subjectivity during the interpretation process. The polar source probability tomography has shown its full capacity to single out the most probable location of the core of the source bodies.

In the forthcoming formulation, dipolar sources are also postulated and are assumed as the most elementary physical sources that may explain, with proper spatial combinations, the response due to sharp macroscopic discontinuities. Dipolar source probability tomography is expected to give insight into the spatial extent of a source body by evidencing the most probable location of its boundaries. The joint analysis of pole and dipole tomographies appears to be a goal of great importance for a better definition of the sources of anomalies.

## THE PROBABILITY TOMOGRAPHY GENERALIZED APPROACH

Consider a reference coordinate system with the (*x,y*)-plane placed at sea level and the *z*-axis positive downwards, and a 3D datum domain *V* as drawn in figure 1. In particular, the top surface *S* can represent a non-flat ground survey area defined by a topographic function $z_t(x,y)$ and the bottom surface $z_b(x,y)$ can correspond to the maximum depths at which datum points are placed. Let $\mathbf{A}(\mathbf{r})$ be a vector anomaly field at a set of datum points $\mathbf{r} \equiv (x,y,z)$, with $\mathbf{r} \in V$.

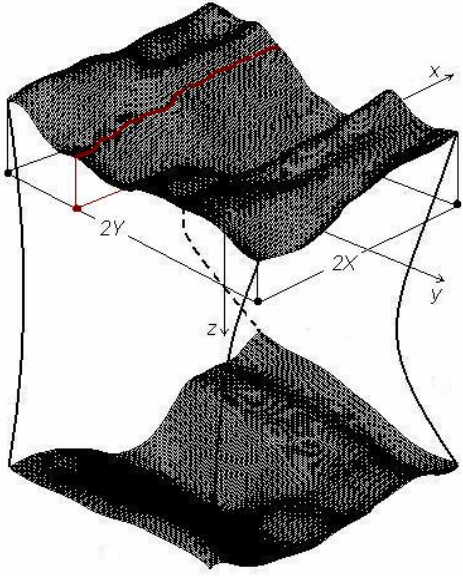

**Figure 1** The 3D datum domain, characterized by irregular boundary surfaces. The $(x,y)$-plane is placed at sea level and the $z$-axis points into the earth.

Assume that $\mathbf{A}(\mathbf{r})$ can be discretized as

$$\mathbf{A}(\mathbf{r}) = \sum_{m=1}^{M} p_m \mathbf{s}(\mathbf{r},\mathbf{r}_m) + \sum_{n=1}^{N} (\mathbf{d}_n \cdot \nabla_n)\mathbf{s}(\mathbf{r},\mathbf{r}_n) , \qquad (1)$$

*i.e.* a sum of effects due to a group of $M$ source poles, whose $m$-th element is located at $\mathbf{r}_m \equiv (x_m,y_m,z_m)$ and has strength $p_m$, and a group of $N$ source dipoles, whose $n$-th element is located at $\mathbf{r}_n \equiv (x_n,y_n,z_n)$ and has as strength the operator $\mathbf{d}_n \cdot \nabla_n$, where $\mathbf{d}_n$ is the dipole moment.

The effect of both the $M$ and $N$ source elements at a datum point $\mathbf{r} \in V$ is analytically described by the same vector kernel $\mathbf{s}(\mathbf{r})$.

We define the information power $\Lambda$, associated with $\mathbf{A}(\mathbf{r})$, within the volume $V$ as

$$\Lambda = \int_V \mathbf{A}(\mathbf{r}) \cdot \mathbf{A}(\mathbf{r}) dV , \qquad (2)$$

which, using eq.1, is expanded as

$$\Lambda = \sum_{m=1}^{M} p_m \int_V \mathbf{A}(\mathbf{r}) \cdot \mathbf{s}(\mathbf{r},\mathbf{r}_m) dV + \sum_{n=1}^{N} \sum_{\nu=x,y,z} d_{n\nu} \int_V \mathbf{A}(\mathbf{r}) \cdot \frac{\partial \mathbf{s}(\mathbf{r},\mathbf{r}_n)}{\partial \nu_n} dV . \qquad (3)$$

*The source pole occurrence probability*

We consider a generic $m$-th integral of the first sum in eq.3 and apply Schwarz's inequality, obtaining

$$\left[ \int_V \mathbf{A}(\mathbf{r}) \cdot \mathbf{s}(\mathbf{r},\mathbf{r}_m) dV \right]^2 \leq \int_V A^2(\mathbf{r}) dV \int_V s^2(\mathbf{r},\mathbf{r}_m) dV , \qquad (4)$$

where $A(\mathbf{r})$ and $s(\mathbf{r},\mathbf{r}_m)$ are the modulus of $\mathbf{A}(\mathbf{r})$ and $\mathbf{s}(\mathbf{r},\mathbf{r}_m)$, respectively.

From inequality 4 we can readily deduce a *source pole occurrence probability* (SPOP) function as

$$\eta_m^{(P)} = C_m \int_V \mathbf{A}(\mathbf{r}) \cdot \mathbf{s}(\mathbf{r},\mathbf{r}_m) dV , \qquad (5)$$

where

$$C_m = \left[ \int_V A^2(\mathbf{r}) dV \int_V s^2(\mathbf{r},\mathbf{r}_m) dV \right]^{-1/2} . \qquad (6)$$

The 3D SPOP function satisfies the condition

$$-1 \leq \eta_m^{(P)} \leq +1 \qquad (7)$$

and is assumed to give a measure of the probability which a source pole of strength $p_m$ placed at $\mathbf{r}_m$ obtain as responsible of the observed anomaly field $\mathbf{A}(\mathbf{r})$.

In some geophysical applications, the anomaly field and kernel functions can directly be described by scalar functions. In this case, it is very easy to show that $\eta_m^{(P)}$ can be calculated using the expression

$$\eta_m^{(P)} = C_m \int_V A(\mathbf{r}) s(\mathbf{r},\mathbf{r}_m) dV , \qquad (5a)$$

using again eq.6 to obtain $C_m$.

The concept of *probability* ascribed to the function $\eta_m^{(P)}$ is motivated as follows. In general, a probability measure $\Psi$ is defined as a function assigning to every subset $E$ of a space of states $U$ a real number $\Psi(E)$ such that (Gnedenko 1979)

$$\Psi(E) \geq 0, \text{ for every } E, \qquad (8a)$$
$$\text{if } E \cap F \equiv 0, \text{ with } E,F \subset U, \Psi(E \cup F) = \Psi(E) + \Psi(F), \qquad (8b)$$
$$\Psi(U) = 1. \qquad (8c)$$

Considering that the presence of a source pole $p_m$ at $\mathbf{r}_m$ is independent from the presence of another source pole at another point, the function

$$\Psi_m = \frac{\left| \eta_m^{(P)} \right|}{\int_{V_m} \left| \eta_m^{(P)} \right| dV_m} , \qquad (9)$$

can be defined as a probability density, since it allows a measure of the probability to find $p_m$ at $\mathbf{r}_m$ to be got in agreement with axioms 8a, 8b and 8c.

Practically, $\eta_m^{(P)}$ is different from $\Psi_m$ only for an unknown constant factor and has also the advantage to give the sign of the source. Thus we can conventionally assume $\eta_m^{(P)}$ as the probability measure of source pole occurrence at $\mathbf{r}_m$.

The $\eta_m^{(P)}$ function, specified in eq.5, can readily be calculated knowing the mathematical expression of the function $\mathbf{s}(\mathbf{r},\mathbf{r}_m)$, which is given the role of *source pole elementary scanner* (SPES) in the SPOP tomographic procedure we are going to explain.

The SPOP 3D tomography of a geophysical dataset consists in a scanning procedure operated by the SPES function. In practice, as the distribution of the source poles responsible of an anomaly field $\mathbf{A}(\mathbf{r})$ is unknown, we place a virtual source pole of unitary strength at the nodes of a grid filling a supposed target space beneath the ground surface. Since the elementary kernel $\mathbf{s}(\mathbf{r})$ is a known function, although assuming different forms according to the geophysical method that is considered, we can compute the SPOP value at each node of the grid, using a digitized form of the integral in eq.5. A positive SPOP value gives the probability with which a positive pole, located where the SPOP value has been computed, can be considered responsible of the $\mathbf{A}(\mathbf{r})$ field. Conversely, a negative value gives the occurrence probability of a negative source pole.

We show now a very simple example in order to let the main qualitative aspects of the SPOP tomography be soon clarified. Figure 2 shows two uniform spheres immersed in a homogeneous half-space at a depth and horizontal distance as displayed in the top section. The contrasts between the constitutive physical parameter of the two spheres and that of the hosting medium are of opposite sign. The anomaly field is supposed to be depicted by a scalar quantity, *e.g.* the *z*-component of the gravitational field, and the corresponding dataset to consist of measures taken on the ground. The anomaly field map is shown, qualitatively, in the mid horizontal slice in figure 2.

As the 3D *V*-domain, previously introduced, in this case collapses to a 2D *S*-domain, the volume integrals appearing in the previous equations reduce to surface integrals extended over *S*, which in general is a non-flat portion of the ground surface. The SPOP $\eta_m^{(P)}$ function is now written as

$$\eta_m^{(P)} = C_m \int_S A(\mathbf{r}) s(\mathbf{r},\mathbf{r}_m) dS ,  \qquad (10)$$

where

$$C_m = \left[ \int_S A^2(\mathbf{r}) dS \int_S s^2(\mathbf{r},\mathbf{r}_m) dS \right]^{-\frac{1}{2}} . \qquad (11)$$

Assuming the projection of *S* onto the $(x,y)$-plane can be fitted to a rectangle *R* of sides 2*X* and 2*Y* along the *x*- and *y*-axis, respectively, as in figure 1, using a topography surface regularization factor $g(z)$ given by

$$g(z) = \left[ 1 + (\partial z/\partial x)^2 + (\partial z/\partial y)^2 \right]^{\frac{1}{2}} , \qquad (12)$$

eq.10 and eq.11 can be regularized as

$$\eta_m^{(P)} = C_m \int_R A(\mathbf{r}) s(\mathbf{r},\mathbf{r}_m) g(z) dxdy \qquad (13)$$

and

$$C_m = \left[ \int_R A^2(\mathbf{r}) g(z) dxdy \int_R s^2(\mathbf{r},\mathbf{r}_m) g(z) dxdy \right]^{-\frac{1}{2}} \qquad (14)$$

where the integration intervals on the *x*-axis and *y*-axis are [-*X*,*X*] and [-*Y*,*Y*], respectively.

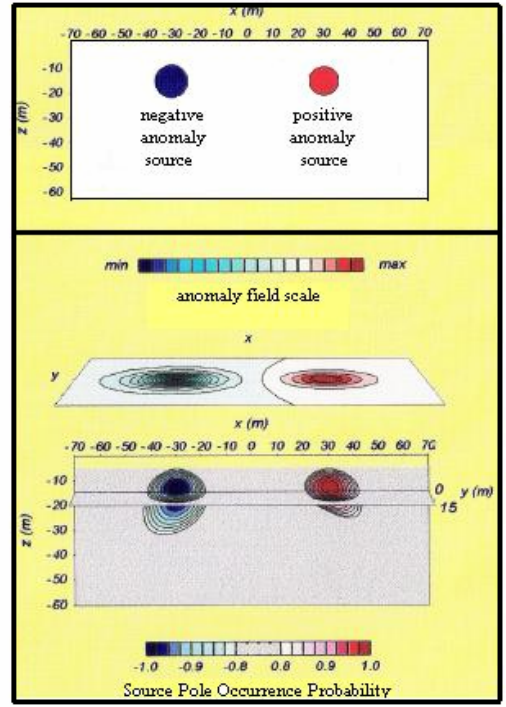

**Figure 2** A qualitative example of the SPOP tomography. The two-sphere model generating anomalies of opposite sign.

For the example of figure 2, *S* has been taken a flat horizontal surface, hence from eq.12 it follows $g(z)=1$.

The SPOP tomography algorithm is applied with the aim of designing a probability space of occurrence of elementary source poles, responsible of the detected anomalies on the ground surface. Therefore, in order to enhance the filtering property of the scanner procedure, in making the integrations in eq.13 and eq.14 it can be advisable to select different integration surfaces, letting *X* and *Y* vary from sufficiently small values up to the size of the survey area. The smallest values for *X* and *Y* can be chosen such that the anomaly generated by a source element located at a node of the grid is entirely contained in the surface domain [-*X*,*X*]×[-*Y*,*Y*] centered over the source pole. At each node, the computations must be carried out for all sizes of the (*X*,*Y*)-domain. The highest calculated value will be assumed with its sign as the most correct SPOP value for that point.

The result of the SPOP tomography applied to the two-sphere example is depicted in the lower section of figure 2. A diffuse picture of the most probable source poles location is the peculiar feature of this new type of target detection approach. The vertical and horizontal positions of the barycentres of the spheres is exactly reconstructed, since they correspond with the points where the highest absolute values of the SPOP function have been got. As expected, no evidence can, however, be deducted as it concerns the size of the spheres.

*The source dipole occurrence probability*

We now consider a generic *n*-th integral of the second double sum in eq.3, apply again Schwarz's inequality and obtain for *ν*=*x*,*y*,*z*

$$\left[\int_V \mathbf{A}(\mathbf{r}) \cdot \frac{\partial \mathbf{s}(\mathbf{r},\mathbf{r}_n)}{\partial \nu_n} dV\right]^2 \leq \int_V A^2(\mathbf{r}) dV \int_V \left|\frac{\partial \mathbf{s}(\mathbf{r},\mathbf{r}_n)}{\partial \nu_n}\right|^2 dV, \quad (15)$$

from which we deduce a 3D *source dipole occurrence probability* (SDOP) function as

$$\eta_{n\nu}^{(D)} = C_{n\nu} \int_V \mathbf{A}(\mathbf{r}) \cdot \frac{\partial \mathbf{s}(\mathbf{r},\mathbf{r}_n)}{\partial \nu_n} dV, \quad (16)$$

where

$$C_{n\nu} = \left[\int_V A^2(\mathbf{r}) dV \cdot \int_V \left|\frac{\partial \mathbf{s}(\mathbf{r},\mathbf{r}_n)}{\partial \nu_n}\right|^2 dV\right]^{-1/2}. \quad (17)$$

Again, for scalar anomaly field and kernel function, $\eta_{n\nu}^{(D)}$ can be easily derived as

$$\eta_{n\nu}^{(D)} = C_{n\nu} \int_V A(\mathbf{r}) \frac{\partial s(\mathbf{r},\mathbf{r}_n)}{\partial \nu_n} dV, \quad (16a)$$

with $C_{n\upsilon}$ given by

$$C_{n\nu} = \left[\int_V A^2(\mathbf{r}) dV \cdot \int_V \left[\frac{\partial s(\mathbf{r},\mathbf{r}_n)}{\partial \nu_n}\right]^2 dV\right]^{-1/2} \quad (17a)$$

Therefore, at each point $\mathbf{r}_n$ three values of $\eta_{n\nu}^{(D)}$ can be calculated. They give a measure of the probability, with which the three components of the source dipole located at $\mathbf{r}_n$ can be considered responsible of the **A**(**r**) field. We expect that maxima of $|\eta_{n\nu}^{(D)}|$ should reveal structural discontinuity surfaces.

Also the SDOP function can be calculated knowing the mathematical expression of the derivatives of the base function **s**(**r**,**r**$_n$). Each derivative is now assigned the role of *source dipole elementary scanner* (SDES) function.

The SDOP 3D tomography of a geophysical dataset consists again in a scanner procedure operated this time by the SDES function. In practice, as we do not know the distribution of the dipoles contributing to generate the anomaly field **A**(**r**), we adopt now a virtual source dipole of unitary strength and put it at the same nodes of the grid, as previously. We can compute the SDOP value at each node by a digitized form of the integral in eq.16. A non-vanishing value of any of the three SDOP functions gives the probability with which the relative component of a source dipole moment, localized at the point where the SDOP value has been computed, can be considered responsible of the given **A**(**r**) field. The algebraic sign indicates now the direction of the dipole component along the axis to which the SDOP function under examination refers.

We show now another synthetic example in order to illustrate the joint peculiarities of the SPOP and SDOP imaging. Figure 3a shows two infinitely extended half-plates in contact, characterized by opposite contrasts of their constitutive parameter with respect to the hosting half-space. The graph of a scalar function, simulating *e.g.* the Bouguer gravity anomaly along a profile *L*, is reported. Since the 3D *V*-domain collapses to a 1D *L*-domain, the volume integrals in all previous SPOP and SDOP equations reduce to line integrals extended over *L*, which, in general, can be a non-straight line.

Assuming that the *L*-domain lies completely on a plane normal to the (*x*,*y*)-plane as the red line in figure 1, and its projection onto this plane, parallel to *x*-axis, is a straight segment of length 2*X*, using a topographic line regularization factor *h*(*x*) expressed by

$$h(x) = \left[1 + (dz/dx)^2\right]^{1/2}, \quad (20)$$

the SPOP $\eta_m^{(P)}$ is expressed by

$$\eta_m^{(P)} = C_m \int_{-X}^{X} A(\mathbf{r}) s(\mathbf{r},\mathbf{r}_m) h(x) dx, \quad (21)$$

where

$$C_m = \left[\int_{-X}^{X} A^2(\mathbf{r}) h(x) dx \int_{-X}^{X} s^2(\mathbf{r},\mathbf{r}_m) h(x) dx\right]^{-1/2}. \quad (22)$$

Accordingly, as only the *x*-derivative of $s(\mathbf{r},\mathbf{r}_n)$ can provide useful results from this 1D simulated curve, the $\eta_{nx}^{(D)}$ regularized integral can be written as

$$\eta_{nx}^{(D)} = C_{nx} \int_{-X}^{X} A(\mathbf{r}) \frac{\partial s(\mathbf{r},\mathbf{r}_n)}{\partial x_n} h(x) dx, \qquad (23)$$

where

$$C_{nx} = \left\{ \int_{-X}^{X} A^2(\mathbf{r}) h(x) dx \cdot \int_{-X}^{X} \left[ \frac{\partial s(\mathbf{r},\mathbf{r}_n)}{\partial x_n} \right]^2 h(x) dx \right\}^{-\frac{1}{2}}. \qquad (24)$$

For the example in figure 3, *L* has been assumed a horizontal straight-line segment, therefore, from eq.20, it follows $h(z)=1$.

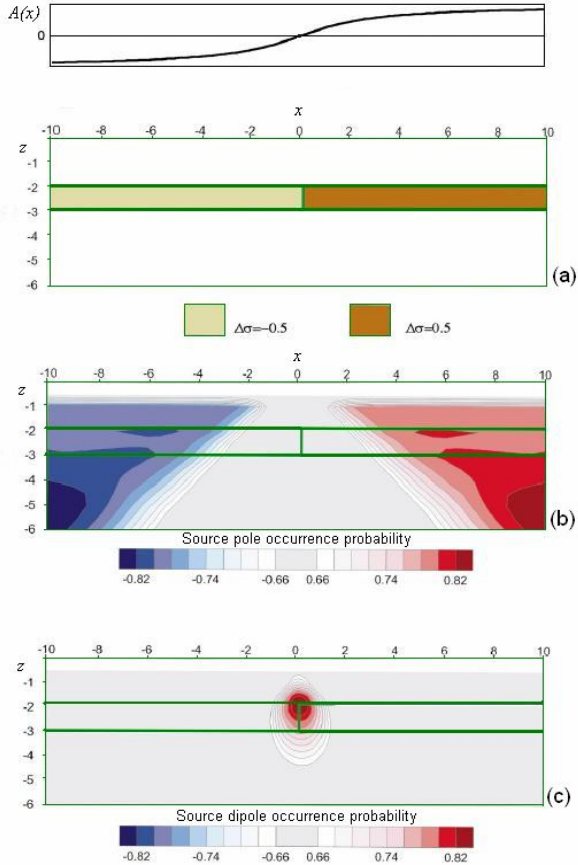

**Figure 3** Application of the SPOP and SDOP tomographies to a simulated prospection along a profile perpendicular to the strike of a 2D structure. (a) Cross-section of the 2D model and relative anomaly profile. (b) SPOP tomography. (c) SDOP tomography.

Figure 3b shows the result of the SPOP tomography along a vertical section through the profile, normal to the strike direction of the 2D target. Two pairs of SPOP nucleuses with opposite sign can be recognized at two different depth levels. The top pair is located inside the half-plates, while the bottom pair is placed well beyond the half-plates with a larger separation than that of the top pair. In each pair the sign of each nucleus conforms to the sign of the constitutive parameter contrast of the corresponding half-plate. The highest SPOP values in modulus belong to the deeper pair, which thus provides an equivalent distributions of source poles compatible with the given anomaly profile. The top pair, to which smaller SPOP values correspond in modulus, appears to correctly outline the top level of the two half-plates. The top pair of poles thus appears to be a geometrically correct equivalent representation of the semi-infinite plates. The comparison between the two pairs of source poles shows that the top poles marking the existence of the true 2D model have less probability to occur than the deeper equivalent source poles, which actually do not correspond to any structure in the starting model.

The SDOP tomography can now be used to try to solve this clear case of equivalence. As previously said, due to the 1D nature of the dataset, from eq.23 only the horizontal *x*-derivative has been calculated. The result of the SDOP tomography is illustrated in figure 3c, wherein only a nucleus very neatly appears astride the top portion of the lateral discontinuity between the two half-plates. It extends down to the bottom wedge of the contact with still appreciable SDOP values. Its positive sign indicates the direction of the horizontal component of the dipole moment, which in full agreement with the assumed model lies along the positive direction of the *x*-axis from left to right. This is a very important result, which confirms our basic assumption of the joint SPOP and SDOP tomography as an efficient approach to the analysis of the most probable equivalent solutions of a given inversion problem.

## CONCLUSION

Adhering to the propensity interpretative approach of modern science (Marshall and Zohar, 1997) a probability tomography method has been developed in order to analyze geophysical vector or scalar field datasets. The purpose of the new method is to get rid of some restrictive approaches that are conventionally adopted to interpret geophysical datasets, usually derived from very simplistic assumptions about the physical reality, often dictated by some more or less idealized or exotic geological concepts. We postulate that the geophysical reality consists of two kinds of reality, say actual and potential, where we mean for actual what we get when we can directly explore the geophysical entity, and for potential the largest spread of possible structural configurations compatible with the measured datasets.

The new method has been tested on a few synthetic examples, which demonstrate that the whole spectrum of the potential solutions to a geophysical interpretation problem can coexist, from a probabilistic point of view, with the actual model, *i.e.* with the model which can then be proved to closely represent the true situation.

As previously said, we have gone through the SPOP tomography again in order to introduce in a direct and more fluent way the innovative argument of the source dipole tomography. For details about the application of the SPOP tomography method to specific geophysical methods, the interested reader is referred to the papers recalled in the introduction.

**Acknowledgements**   Study performed with financial grants from the Italian Ministry of Education, University and Research (PRIN 2000 project), the European Commission (TOMAVE project) and the Italian Group of Volcanology of the National Research Council.